\documentclass[prd,twocolumn,floatfix,amsmath,nofootinbib,amssymb,floatfix,superscriptaddress]{revtex4}
\usepackage{graphicx,color,dcolumn,booktabs,bm}
\usepackage{longtable,lscape}
\usepackage{pdfpages}
\usepackage{txfonts}
\usepackage{overpic}
\usepackage{amssymb}
\usepackage{makecell}
\usepackage{indentfirst}
\usepackage{feynmf}   
\usepackage{slashed}  
\usepackage{cases}
\usepackage{color}
\usepackage{multirow}
\usepackage{threeparttable}
\usepackage{epstopdf}
\usepackage{enumerate}
\usepackage{subfigure}
\usepackage{diagbox}
\usepackage{graphicx,color,dcolumn,booktabs,bm}
\usepackage{mathrsfs}
\usepackage{cancel}
\usepackage{float}
\usepackage[misc]{ifsym}
\usepackage{array}
\usepackage{tikz}
\usetikzlibrary{decorations.markings}
\usepackage[colorlinks,
            citecolor=blue,
            anchorcolor=red,
            menucolor=red,
            linkcolor=red,
            filecolor=red,
            runcolor=red,
            urlcolor=blue,
            frenchlinks=red]{hyperref}
\usepackage{xcolor}

\begin{document}
	\title{Understanding the 1P- and 2S nucleon resonances within the extended Lee-Friedrichs Model}
	
	\author{Yu-Hui Zhou}
\affiliation{School of Physics, Southeast University, Nanjing 211189,
P.~R.~China}
\author{Hui-Hua Zhong}
\affiliation{College of Aeronautics Mechanical and Electrical Engineering, Jiangxi Flight University, Nanchang 330088, China
}
    \author{Zhi-Yong Zhou}
    \email[]{zhouzhy@seu.edu.cn}
\affiliation{School of Physics, Southeast University, Nanjing 211189,
P.~R.~China}

\author{Xian-Hui Zhong}
\email[]{zhongxh@hunnu.edu.cn}
\affiliation{Department of Physics, Hunan Normal University, and Key Laboratory of Low-Dimensional
Quantum Structures and Quantum Control of Ministry of Education, Changsha 410081, China
}
\affiliation{Synergetic Innovation Center for Quantum Eﬀects and Applications (SICQEA), Hunan Normal University, Changsha 410081, China
}

	\date{\today}
	
	\begin{abstract}
		We present a unified desciption of the low-lying $1P$ and $2S$-wave nucleon resonance within the framework of an extended Lee-Friedrichs scheme. By incorporating the coupled-channel dynamics between bare quark-model states and the $\pi N$, $\pi\Delta$ and $\eta N$ meson-baryon continua, we examine the mass shifts and structural properties of these excited states. We demonstrate that when the model parameters are calibrated to match the $1P$-wave spectrum and their widths, the pole associated with the bare $2S$ state is naturally shifted downward to the mass region of physical Roper resonance--$N(1440)$, thereby offering a dynamical explanation for the long-standing level-inversion problem. An approximate analysis of compositeness and elementariness reveals that the Roper resonance contains a significant meson-baryon continuum states, consistent with the picture of a bare core heavily dressed by meson-baryon cloud. Simultaneously, the pole positions and properties of five $1P$-wave resonances--$N(1535)$, $N(1650)$, $N(1520)$, $N(1700)$ and $N(1675)$ are successfully reproduced. Our results highlight the essential role of coupled-channel effects in shaping the nucleon spectrum and provide a consistent microscopic insight into the interplay between internal quark degrees of freedom and external hadronic fields.

	\end{abstract}
	
	\maketitle
	
	\section{Introduction}

Understanding the spectroscopy of excited baryon resonances is crucial for elucidating the non-perturbative dynamics of Quantum Chromodynamics~(QCD) in the confinement regime. In recent decades, high-precision photoproduction and electroproduction experiments at facilities such as JLab, MAMI, ELSA, BES III and Belle II have amassed a wealth of data, significantly enriching the baryon spectrum. However, extracting reliable resonance parameters from these data remains a formidable challenge. Similar to the meson sector or even worse, the baryon resonances are usually broad and overlapping with each other, which makes the identification of their internal structure and properties highly sensitive to the theoretical ansatz employed in the Partial Wave Analysis (PWA). Among these elusive states, the Roper resonance--$N(1440)$, with  spin–parity quantum number $J^P=1/2^+$~\cite{Roper:1964zza}, stands out as one of the most enigmatic puzzles.

The primary theoretical difficulty concerning the Roper resonance is the well-known mass inversion problem. In the conventional quark model based on the one-gluon-exchange (OGE) and confinement interactions, usually the first orbital excited states~($N$=1, $L$=1) are expected to lie significantly lower in mass than the first radial excited states~($N$=2, $L$=0). 
	Indeed, the conventional quark model has been remarkably successful in describing the ``normal" sector of the baryon spectrum; the five experimentally established 
	1P-wave nucleon resonances, such as $ N(1535) 1/2^-$, $N(1520) 3/2^-$, $N(1650) 1/2^-$, $N(1700) 3/2^-$ and $N(1675) 5/2^-$, follow the mass ordering predicted by three-quark dynamics reasonably well. 	However, 	$N(1440)$ appears roughly 100~MeV below the $N(1535)$. This observation implies that the first radial excitation is lighter than the first orbital excitation, a level ordering that directly contradicts the expectations of potential models. Consequently, the nature of the $N(1440)$ has been investigated extensively across various frameworks, including lattice QCD, Faddeev equations and effective field theory~\cite{Suzuki:2009nj,Liu:2016uzk,Qin:2019hgk,Wu:2017qve,Lang:2016hnn,Suenaga:2022ajn,Clement:2020dby,Ball:1967zzb,Zou:2025nnw,Zhao:2006an,Long:2011rt,Chen:2017pse,Tan:2025kjk,Cheng:2025sdp}.

	Motivated by this discrepancy, subsequent efforts has been devoted to refine the consistent quark model formulation.
	The relativized quark model  developed in Refs.~\cite{Godfrey:1985xj,Capstick:1986ter} claimed the problem was tamed when relativistic kinematics are incorporated, yet  these improvements did not change the ordering of the  energy levels.	
	By introducing the one–boson–exchange (OBE) hyperfine interaction~\cite{Glozman:1995fu},  the mass inversion between the Roper resonance and the $N(1535)$ can be reproduced. Unlike the flavor-blind confinement potential, the spin-flavor dependence of the OBE interaction provides additional attraction for the spatially symmetric wave function of the Roper resonance, thereby lowering its mass below that of the $N(1535)$. Furthermore, in hybrid quark models incorporating both OGE and OBE potentials, it was found that the mass of $N(2S,1/2^+)$ is lowered  about 400 MeV, whereas the $1P$-wave nucleons  are only shifted  by only roughly 100 MeV~\cite{PhysRevD.110.116034}.
	Similar conclusions emerge from unquenched quark models~\cite{Julia-Diaz:2006odw}, where the screening effects of quark-antiquark pairs are considered. These results collectively suggest that the Roper resonance cannot be viewed as a pure three-quark state but new mechanisms should be adopted to understand its nature.

Beyond the mass spectrum, the exceptionally large width of the Roper resonance also poses a severe challenge for physical interpretation~\cite{ParticleDataGroup:2024cfk,Bijker:2015gyk,Hunt:2018wqz,Gegelia:2016xcw,Burkert:2017djo,Shklyar:2012js}. The commonly used Breit-Wigner formulism, while adequate for narrow states, becomes unreliable for such broad resonances since the threshold effects and unitarity constraints usually play an important role in correctly understanding their information. Rigorously, a resonance is defined as a pole of the $S$-matrix on the un-physical Riemann sheet of complex energy plane. To understand the dynamical origin of these broad resonances, it is necessary to  consider the coupled-channel effects  and	study the properties by analytically continuing  into the complex energy plane.

	To date, several studies untilizing the dynamical coupled-channel models have made progress in extracting the Roper pole from $\pi N$ and $\pi\pi N$ scattering data, providing valuable insights into multi-channel resonance dynamics~\cite{Hoferichter:2023mgy,Ronchen:2022hqk,Hunt:2018wqz,Anisovich:2011fc,Arndt:2006bf,Pearce:1990uj,Suzuki:2009nj,Wang:2023snv}.  Suzuki et al.~\cite{Suzuki:2009nj} claimed that $N(1440) $ and $N(1710)$  evolve from a single bare nucleon state with a mass of 1.763 GeV by tracing pole trajectories as the couplings were turned on. In a different approach, Wang et al.~\cite{Wang:2023snv}
	demonstrated that $N(1440)$ can also be generated purely dynamically through $t$- and $u$-channel hadron-exchange dynamics, without introducing any $s$-channel bare state. Their result is consistent with early results of a coupled-channel meson-exchange model~\cite{Krehl:1999km}. These varying interpretations highlight the need for a theoretical framework that can transparently disentangle the contributions of intrinsic quark core from the dynamical dressing of the meson cloud.
	
	Along these lines, we employ the extended Lee-Friedrichs~(LF) scheme~\cite{Xiao:2015gra,Xiao:2016mon,Xiao:2016dsx,Xiao:2023lpv}. This formalism offers a rigorous solvable Hamiltonian framework that allows for an exact treatment of the coupling between discrete bare states and continuum channels, fully preserving unitairty and analyticity. 
	The exact solvability provides a clear and transparent way to disentangle the contributions of bare
 discrete state and coupled continuum states, which could be applied into the baryon systems to study the  intrinsic quark-core and coupled channels.  Furthermore, embedding this scheme into the rigged Hilbert space~(RHS)  providing a solid mathematical foundation for understanding the complex eigenvalue of Hamiltonian, their wave functions, and other related properties~\cite{Gadella:2004}.
	Consequently, this approach offers a complementary perspective on the formation and properties of the resonance, linking underlying quark dynamics directly to observable features such as pole positions and widths.

In this work, we  extend the LF model combined with the quark pair creation (QPC) model to the nucleon system for the first time, aiming to clarify the dynamical origin and structure of the Roper resonance and other orbital excited state through a consistent coupled channel analysis.
	
	This paper is organized as follows. The theoretical foundations are introduced in Sec.~\ref{sec:theory}, which include the Lee-Friedrichs model, the potential model and the QPC model. The numerical calculations and analysis are given in Sec.~\ref{sec:results}. Sec.~\ref{Summary} contains our final conclusion and summary.
	
	\section{theoretical framework}
	\label{sec:theory}
	\subsection{ The extended Lee-Friedrichs scheme}
	\label{Leemodel}
	
	The Lee-Friedrichs model describes how resonance phenomena occur when a discrete bare state couples to a continuum state~\cite{Lee:1954iq,Friedrichs:1948}. When extended to a more general version with multiple continuum states and with multiple discrete states, the model remains to  be exactly solvable and satisfies the unitarity of the $S$-matrix so that it has a more general phenomenological application in   hadron physics~\cite{Xiao:2015gra,Xiao:2016dsx,Xiao:2016mon,Xiao:2023lpv}. When we consider a discrete bare state coupled with multiple continuum states~(or called scattering channels), the  total Hamiltonian is given by $H = H_0 + V$, where  $H_{0}$  represents the free Hamiltonian and  $V$ the interaction term. The free Hamiltonian is expressed as
	\begin{equation}
		H_0 = m_0 |1\rangle \langle 1| +\sum_{i} \int_{\omega_{{\rm th,i}}}^{\infty}E|E,i\rangle\langle E,i|dE,
	\end{equation}
where $|1\rangle$ and $|E ,i\rangle $  denote the discrete  state and the $i$-th continuum state, respectively. 
{The normalization conditions for the bare states are
\begin{equation}
\begin{aligned}
    & \langle1|1\rangle=1,\,\,\langle 1|E,i\rangle=\langle E,i|1\rangle=0,\\
    &\langle E,i|E^{'},j\rangle=\delta_{ij}\delta(E-E^{'}).
\end{aligned}  
\end{equation}
}
The index $i$ labels the quantum numbers of a given continuum state, including the particle species,  total spin $S$ and orbital angular momentum $L$.  We denote the bare mass of the discrete state by   $m_{0}$  while   $\omega_{{\rm th,i}}$ represents the energy threshold of  the $i$-th continuum state. The interaction term is expressed as
\begin{equation}
		V = \sum_{i} \int_{\omega_{{\rm th,i}}}^{\infty}\left( f_{i}(E,{i})|E,{i}\rangle \langle 1|+f^{*}_{i}(E)|1\rangle \langle E,{i}|\right)dE,
	\end{equation}
where	the coupling between $|1\rangle$ and $|E_{i}\rangle$ is characterized by a vertex function $f_{i}(E)$ . The conservation of the total angular momentum and  its third component is implicit.  Solving the eigenvalue problem 
\begin{equation}
		H |\Psi(x)\rangle = x |\Psi(x)\rangle,
		\label{eq:eigenvalue}
	\end{equation}
within the RHS provide a solid mathematical foundation for studying resonance phenomenon in the coupled system. 
{
Because of the completeness relation
	\begin{equation}
	|1\rangle\langle 1| +\sum_{i} \int_{\omega_{\mathrm{th,i}}}^{\infty} |E,i \rangle\langle E,i|\, dE = 1,	 
	\end{equation}}
the general solution of wave function $\Psi(x)$ is supposed to be expanded in the basis of discrete and continuum states as
	\begin{equation}
		|\Psi(x)\rangle = \alpha(x) |1\rangle +\sum_{i} \int_{\omega_{{\rm th,i}}}^\infty \psi_i(x,E)|E,{i}\rangle dE,
	\end{equation}
	where $\alpha(x)$ is a complex function and $\psi_{i}(x,E)$ represents a distribution over the continuum states.
By substituting the general solution to the eigenfunction and projecting it on the bases, the solutions could be written down as
	 \begin{align}
     |\Psi_{0}(x)\rangle& =\alpha(x)\left[|1\rangle + \sum_i \int_{\omega_{{\rm th,i}}}^{\infty} \frac{f_i(E)}{x - E + i\epsilon} |E,i\rangle \, dE\right],\  (x<\omega_{{\rm th,1}})\nonumber\\
	|\Psi_{i}^\pm(x)\rangle &= |x\rangle_i + \frac{f_i^*(x)}{\eta^\pm(x)} \left[|1\rangle + \sum_i \int_{\omega_{{\rm th,i}}}^{\infty} \frac{f_i(E)}{x - E + i\epsilon} |E,i\rangle \, dE\right],\nonumber\\ 
  &  \ \ \ \ \ \ \ \ \ \ \ \ \ \ \ \ \ \ \ \ \ \ \ \ \ \ \ \ \ \ \ \ \ \ \ \ \ \ \ \ \ \ \ \ \ \ (\omega_{{\rm th,i}}<x<\omega_{\rm th,i+1})
	\end{align}
where  $i\epsilon$	in the denominator is introduced for avoiding the singularity at $x=E$. From the prospective of scattering theory, $|\Psi_{i}^+(x)\rangle$ is the ``in" state and $|\Psi_{i}^-(x)\rangle$ is the ``out" state. The scattering $S$ matrix is obtained by the inner product of the ``in" and ``out" state, and the unstable intermediate state is represented by the pole of the scattering amplitude. The key ingredient in solving the eigenvalue problem of Eq.~(\ref{eq:eigenvalue}), determining the resonance eigenvalues and constructing the wave functions within the RHS is the inverse resolvent function $\eta(x)$, expressed as  
\begin{equation}
\eta(x) ^\pm= x - m_0 - \sum_{i} \int_{\omega_{{\rm th,i}}}^{\infty} \frac{|f_i(E)|^2}{x - E\pm i\epsilon } \, dE.	
\label{etax}
\end{equation}
The $\eta(x) ^\pm$ function has  cuts  starting from every energy threshold $\omega_{{\rm th,i}}$ to the infinity. After analytical continuation to different Riemann sheets~(RS) on the complex energy plane, the resonance poles are represented by $\eta(z)=0$. Bound state poles are located below the first threshold on the first RS while resonances  are located on the other RSs.  
The resonance pole is defined as
\begin{equation}
	z_{R}=M-\dfrac{i}{2}\Gamma,
\end{equation}
 where $M$ and  $\Gamma$ are the mass and  width of the resonance pole, respectively.

For multi-channel case, each threshold $\omega_{\rm th,i}$ introduces a branch cut, resulting in a total of $2^n$ Riemann sheets.
In this paper, we are going to deal with the situation with three  channels,
\begin{align}
	\eta^{\mathrm{II}}(z) &= \eta^{\mathrm{I}}(z) - 2 \pi i G_1(z), & \label{eq:etaII} \\
	\eta^{\mathrm{III}}(z) &= \eta^{\mathrm{I}}(z) - 2 \pi i G_1(z) - 2\pi i G_2(z), & \label{eq:etaIII} \\
	\eta^{\mathrm{IV}}(z) &= \eta^{\mathrm{I}}(z)- 2 \pi i G_1(z) - 2 \pi i G_2(z)-2\pi i G_{3}(z), & \label{eq:etaIV}
\end{align}
in which  $G_i(z)=f_i(z)f_i^{*}(z)$ represents the analytical continuation of the module square of the coupling function of the $i$-th channel and I, II, III and IV label the first, second, third and fourth sheets, respectively.
Here, we only focus on the un-physical Riemann sheets which is closest to the physical sheet, since the poles on these sheets  are close to the physical region.

 The wave function of a bound state with real eigenvalue $z_{B}$ can be explicitly represented as
 \begin{align}\label{boundstatewavefunction}
 	|z_{B}\rangle&=N_{B}\bigg(|1\rangle +\sum_{i}\int_{\omega_{{\rm th},i}}^{\infty}\dfrac{f_{i}(E)}{z_{B}-E}dE|E,{i}\rangle\bigg),
 \end{align} with the normalization defined as $\langle  z_B|z_B\rangle=1$, and \begin{equation}
	N_{B}=\bigg(1+\sum_{i}\int_{\omega_{{\rm th},i}}^{\infty}\dfrac{|f_i(E)|^2}{(z_{B}-E)^{2}}dE\bigg)^{-1/2}.
\end{equation}

 The wave function of a resonance with complex eigenvalue $z_{R}$ can be explicitly represented as
  \begin{equation}
  \begin{aligned}
 	|z_{R}\rangle&=N_{R}\bigg(|1\rangle +\sum_{i}\int_{\omega_{{\rm th},i}}^{\infty}\dfrac{f_{i}(E)}{[z_{R}-E]_{+}}dE|E,{i}\rangle\bigg),\nonumber\\
 	\langle \tilde z_{R}|&=N_{R}\bigg(\langle 1| +\sum_{i}\int_{\omega_{{\rm th},i}}^{\infty}\dfrac{f_{i}(E)}{[z_{R}-E]_{+}}dE\langle E,{i}|\bigg).
 \end{aligned}
 \label{resonancewavefunction}   
 \end{equation}

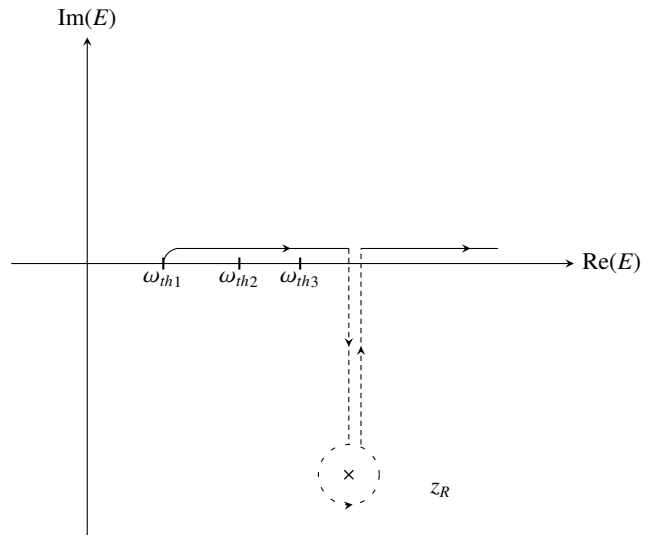
\begin{figure}
    \centering
 \begin{tikzpicture}[scale=2]

		\draw[->, >=stealth](-0.2,0) -- (3.5,0)
		node[right] {  $\mathrm{Re}(E)$};
		
		\draw[->, >=stealth](0.3,-1.8 ) -- (0.3,1.5)
		node[above] {  $\mathrm{Im}(E)$};
		
\draw[line width= 0.8pt]
(0.8,-0.04) -- (0.8,0.04)
node[below, yshift=-2pt] { \small ${\omega_{th}}_{1}$};

\draw[line width=0.8pt]
(1.7,-0.04) -- (1.7,0.04)
node[below, yshift=-2pt] { \small ${\omega_{th}}_{3}$};

\draw[line width=0.8pt]
(1.3,-0.04) -- (1.3,0.04)
node[below, yshift=-2pt] { \small ${\omega_{th}}_{2}$};

\draw[line width=0.4pt,
postaction={decorate},
decoration={
	markings,
	mark= at position 0.7 with  {\arrow[>=stealth]{>}}
}]
(0.8,0)
.. controls (0.8,0.05) and (0.85,0.1) .. (0.9,0.1) (0.9,0.1)   
-- (2.02,0.1);
\draw[line width=0.4pt,
postaction={decorate},
decoration={
	markings,
	mark= at position 0.8 with {\arrow[>=stealth]{>}}
}]
(2.1,0.1) -- (3,0.1);

	\draw[dash pattern=on 2pt off 1.5pt,
	postaction={decorate},
	decoration={
		markings,
		mark= at position 0.5 with {\arrow[>=stealth]{>}}
	}]
	(2.02,0.1) -- (2.02,-1.2);
	
	\draw[dash pattern=on 2pt off 1.5pt,
	postaction={decorate},
	decoration={
		markings,
		mark= at position 0.5 with {\arrow[>=stealth]{>}}
	}]
	(2.1,-1.2) -- (2.1,0.1);

\draw[
dash pattern=on 1.8pt off 4pt,
postaction={decorate},
decoration={
	markings,
	mark=at position 0.77 with {\arrow[>=stealth]{>}} 
}
] 
(2.22,-1.4) arc [start angle=0, end angle=360, radius=0.2];

\draw[line width=0.5pt] (1.99,-1.43) -- (2.05,-1.37);
\draw[line width=0.5pt] (1.99,-1.37) -- (2.05,-1.43);
		\node[right] at (2.5,-1.5 ) {$z_{R}$};	
	\end{tikzpicture}
    \caption{The deformation of the integral path. The dashed line means the path on the RS where the pole is located.}
    \label{path}
\end{figure}
It is worth emphasizing that the subscript ``$+$'' denotes the integration contour which begins at the threshold along the upper rim of the branch cut and it has to be deformed to encircle the resonance pole in a counter-clockwise direction when it passes the physical region  where the RS of resonance pole is attached, as shown in Fig.~\ref{path}~\cite{Xiao:2016dsx}.
The  normalization factor $N_{R}$ is given by 
\begin{equation}
	N_{R}=\bigg(1+\sum_{i}\int_{\omega_{{\rm th},i}}^{\infty}\dfrac{|f_i(E)|^2}{[z_{R}-E]_{+}^{2}}dE\bigg)^{-1/2},
\end{equation}
where the integration contour is similarly defined as mentioned above. The normalization is usually defined as $\langle\tilde z_R|z_R\rangle=1$, since the usually normalization gives $\langle z_R|z_R\rangle=0$.

	\subsection{The Potential Model}
	\subsubsection{Hamiltonian}
	In this framework, the bare mass $m_{0}$  and the  wave function of a nucleon in Eq.~(\ref{etax}) are described by the potential model.
	This treatment is justified as an effective description at low energy regions, where the baryon dynamics are well captured by constituent quarks interacting through phenomenological potentials. While  the  non-relativistic  quark model (NRQM) involves certain simplifications, it has proven successful in describing the light-baryon spectrum, reproducing the SU(3) octet–decuplet structure, hyperfine splittings, and mass systematics across the light-baryon 
	families \cite{Isgur:1978xj,Isgur:1978wd,Bhaduri:1981pn,Kalman:1988br}.  Consequently, it provides a reasonable and widely adopted basis for specifying the bare mass of low-lying excited nucleons in the coupled-channel analysis.  Accordingly, we utilize the NRQM to calculate the bare masses and adopt a Hamiltonian of the form:
	\begin{equation}
		H_{\rm{NRQM}} = \sum_{i=1}^{3} \left( m_i + T_i \right) - T_G \;+\; \sum_{i<j} V_{ij}(r_{ij}) \;+\; C_0 ,
	\end{equation}
	where $m_{i}$ and $T_{i}$  denote the constituent quark mass and kinetic energy of the $i$-th quark, respectively.
	$T_{G}$ is the center-of-mass kinetic energy and $C_{0}$ is the zero point energy. The term
    $V_{ij}$ represents the effective potential between the $i$-th and $j$-th quarks with a distance $r_{ij}\equiv|\mathbf{r}_{i}-\mathbf{r}_{j}|$. 	
	The potential incorporates both the confinement interaction and the OGE contribution, which can be decomposed into spin–independent and spin–dependent components:	\begin{equation}
		 V_{ij}(r_{ij}) = V_{ij}^{\text{si}}(r_{ij}) + V_{ij}^{\text{sd}}(r_{ij}) .
	\end{equation}
	The spin-independent term $V_{ij}^{\text{si}}$  is composed of  a linear and Coulomb potential:
\begin{equation}
V_{ij}^{\text{si}}(r_{ij}) \;=\; \frac{b}{2}\, r_{ij} \;-\; \frac{2}{3}\,\frac{\alpha_s}{r_{ij}} ,	 
\end{equation}
where the  parameter $b$ represents the
strength of the confinement and $\alpha_s$  the strong coupling constant. 
The spin-dependent term comprises spin-spin, spin-orbit and tensor potentials:
\begin{equation}
	V^{\rm sd}_{ij} = V^{\rm SS}_{ij}+ V^{\rm LS}_{ij}  + V^{\rm T}_{ij} .
\end{equation}
	For the spin-spin and tensor terms, we adopt the  widely  used forms:
	\begin{equation}
		V^{SS}_{ij}
		= -\,\frac{2\alpha_s}{3}
		\left\{
		-\frac{\pi}{2}\,
		\frac{\sigma_{ij}^{3} e^{-\sigma_{ij}^{2} r_{ij}^{2}}}{\pi^{3/2}}
		\cdot
		\frac{16}{3\, m_i m_j}
		\left( \mathbf{S}_i \cdot \mathbf{S}_j \right)
		\right\},
	\end{equation}
	\begin{equation}
	V^{T}_{ij}
	=
	\frac{2\alpha_s}{3}\,
	\frac{1}{m_i m_j\, r_{ij}^{3}}
	\left\{
	3
	\frac{
		\left( \mathbf{S}_i \cdot \mathbf{r}_{ij} \right)
		\left( \mathbf{S}_j \cdot \mathbf{r}_{ij} \right)
	}{r_{ij}^{2}}
	-
	\mathbf{S}_i \cdot \mathbf{S}_j
	\right\} ,	 
	\end{equation}
	where	$\mathbf{S}_{i}$ is the spin operator of the $i$-th quark, and $\sigma_{ij}$	is the smearing parameter introduced to regulate the short-range contact interaction.
	For the spin–orbit interaction, we employ the simplified phenomenological form used in Ref.~\cite{Liu:2019wdr} (originally proposed in Refs.~\cite{Pervin:2007wa,Roberts:2007ni}), which is 
	\begin{equation}
		V^{LS}_{ij} = \frac{\alpha_{\mathrm{SO}}}{\rho^{2} + \lambda^{2}}
		 \cdot
		\frac{\mathbf{L} \cdot \mathbf{S}}{3\,(m_1 + m_2 + m_3)^{2}},
	\end{equation}
	where $\mathbf{L}$ and $\mathbf{S}$ correspond to the total orbital and total spin angular momentum of the baryon respectively, and $\alpha_{\mathrm{SO}}$
	is the spin–orbit coupling strength.
	
	\subsubsection{Numerical method}

		The bare masses and spatial wave functions could be obtained by numerically solving the Schrödinger equation with the Hamiltonian above.
	For such a three-body system, the total spatial wavefunction $\psi_{NLM_{L}}(\boldsymbol{\rho}, \boldsymbol{\lambda})$ can be expanded as a combination of  $\psi_{n_{\rho}l_{\rho}m_{\rho}}(\boldsymbol{\rho})$  and $\psi_{n_{\lambda}l_{\lambda}m_{\lambda}}( \boldsymbol{\lambda})$,
\begin{equation}
	\Psi_{N L M_L}(\boldsymbol{\rho}, \boldsymbol{\lambda})
	=\!\! \sum_{\substack{
			\scalebox{0.5}{$N = 2(n_\rho + n_\lambda)$} \\ \scalebox{0.5}{$+ l_\rho + l_\lambda$} \\
			\scalebox{0.5}{$L = l_\rho + l_\lambda$}
	}}\!\!
	C^{\scalebox{0.5}{$n_\rho l_\rho m_\rho$}}_{\scalebox{0.5}{$n_{\lambda}l_{\lambda}m_{\lambda}$}}
\,[\psi_{n_\rho l_\rho m_\rho}(\boldsymbol{\rho}) 
	\psi_{n_\lambda l_\lambda m_\lambda}(\boldsymbol{\lambda})]_{\scalebox{0.7}{$NLM_{L}$}},
\end{equation}
 where $\boldsymbol{\rho}$ describes the relative motion between two quarks, and
 $\boldsymbol{\lambda}$ the motion of the third quark relative to the quark-pair center of mass.
 The $n_\rho$ and $n_\lambda$ are the principal quantum numbers of the
$\rho$- and $\lambda$-mode oscillators, respectively, while the $N,L$ and $M$ are the total principal quantum number, the total orbital quantum number and its $z$ component, respectively.
The $\rho$- and $\lambda$-mode spatial wave functions share the common decomposition:
\begin{equation}
	\psi_{n_\xi l_\xi m_\xi}(\boldsymbol{\xi})
	= R_{n_\xi l_\xi}(\xi) Y_{l_\xi m_\xi}(\hat{\xi}) ,
\end{equation}
with the radial part  $R_{n_\xi l_\xi}(\xi)$ and the angular dependence  encoded in the spherical harmonic $Y_{l_\xi m_\xi}(\hat{\xi})$.
In this work, we employ the multi-Gaussian expansion method (GEM) to calculate the hadron mass spectrum and the radial wave functions. Unlike the conventional Gaussian expansion method proposed by Hiyama et al.~\cite{Hiyama:2003cu}, our trial radial wave functions $R_{n_\xi l_\xi}(\xi)$  are expanded with a series of harmonic oscillator functions:

\begin{equation}
	R_{n_\xi l_\xi}(\xi)
	= \sum_{l=1}^{n}
	C_{\xi\ell}\,\phi_{n_\xi l_\xi} (d_{\xi\ell} , \xi) ,
\end{equation}
where 
\begin{equation}
	\scalebox{0.9}{$
		\begin{aligned}
			\phi_{n_\xi l_\xi}(d_{\xi \ell};\, \xi)
			= & \bigg(\frac{1}{d_{\xi \ell}}\bigg)^{\frac{3}{2}} 
			\left[
			\frac{
				2^{l_\xi +2 - n_\xi} (2l_\xi + 2n_\xi + 1)!! 
			}{\sqrt{\pi} n_\xi!\left[(2l_\xi+1)!!\right]^2}
			\right]^{\frac{1}{2}} \bigg(\frac{\xi}{d_{\xi \ell}}\bigg) \\
			& \times e^{-\frac12 \left(\frac{\xi}{d_{\xi \ell}}\right)^2}
			F\bigg(-n_\xi,\, l_\xi + 3/2,\, \bigg(\frac{\xi}{d_{\xi \ell}}\bigg)^2\bigg),
		\end{aligned}
		$}
\end{equation}
where $F(-n_\xi,\, l_\xi+3/2,\,(\xi/d_{\xi\ell})^2)$ is the confluent hyper-geometric function.
The variational parameter $d_{\xi\ell}$ is related to the harmonic oscillator frequency $\omega_{\xi\ell}$ via
\begin{equation}
	\frac{1}{d_{\xi\ell}^{\,2}} = M_\xi\, \omega_{\xi\ell}. 
\end{equation}
The reduced masses are defined by
\begin{equation}
	M_\rho \equiv \frac{2 m_1 m_2}{m_1 + m_2},\qquad
	M_\lambda \equiv \frac{3 (m_1 + m_2)m_3}{2 (m_1 + m_2 + m_3)}.
\end{equation}
On the other hand, the harmonic oscillator frequency $\omega_{\xi \ell}$
can be related to the oscillator stiffness constant $K_l$ through
\begin{equation}
	\omega_{\xi \ell} = \sqrt{\frac{3K_\ell}{M_\xi}} \, .
\end{equation}
Thus, for an $udd$ system, we can  easily obtain the following equation:
\begin{equation}
	d_{\rho \ell} = d_{\lambda \ell} = d_\ell
	= (3 m_u K_\ell)^{-1/4},
\end{equation}
where $m_u$ is the constituent mass of the up quark.

Concerning the determination of the variational parameters, we adopt the procedure of Hiyama et al.~\cite{Hiyama:2003cu}, in which the parameters are selected according to a geometric progression,
\begin{equation}
	d_{\ell}=d_{1}a^{\ell-1}\,\,(l=1,2,...,n).
\end{equation}
 It is found that
when we take $d_{1}$=0.1   fm, $d_n$=2 fm and  $n$=15,  stable solutions for the $N$ baryons could be obtained.
Finally, the Schrödinger equation can be solved by dealing with the generalized eigenvalue problem,
\begin{equation}
	\sum_{\ell = 1}^{n} \sum_{\ell' = 1}^{n} 
	\big( H_{\ell \ell'} - E_{\ell}\, N_{\ell \ell'} \big) C_{\ell'}^{\ell} = 0 ,
\end{equation}
where the Hamiltonian elements between different Gaussian bases and the overlapping matrix element of the Gaussian bases are respectively expressed as
\begin{equation}
	H_{\ell\ell'} \equiv \langle \Psi(d_{\ell'}) | H | \Psi(d_\ell) \rangle, \quad
	N_{\ell\ell'} \equiv \langle \Psi(d_{\ell'}) | \Psi(d_\ell) \rangle.
\end{equation}

	\subsection{ The Quark Pair Creation  Model}
	We adopt the  Quark Pair Creation (QPC) model to describe the  coupling vertices $f_i(E)$ between the bare baryon state and the meson–baryon continuum. These vertices are expressed as  $f_{SL}(E)$ where the subscripts $S$ and $L$ denote the total spin  and relative angular momentum of the scattering channel respectively.  The QPC model has a long history of success in describing the OZI-allowed strong decay process of mesons and baryons~\cite{Micu:1968mk,Carlitz:1970xb,LeYaouanc:1972vsx}, providing analytical forms for the vertex functions.  Motivated by the successful application of the QPC model to the charmonium system in our previous work~\cite{Zhou:2017dwj}, we extend this framework to the nucleon sector, allowing for a straightforward and consistent extraction of the $N^*\rightarrow$ meson–baryon coupling vertices within the extended Lee-Friedrichs scheme.
	
	In the QPC model, the transition operator for the process \(A \to BC\) in the non-relativistic framework is given by	
	\begin{equation}
		\begin{aligned}
		T = &-3 \, \gamma \, \sum_m \langle 1 m; 1 - m | 0 0 \rangle 
		\int d^3 \mathbf{p}_4 \, d^3 \mathbf{p}_5 \, \delta^3(\mathbf{p}_4 + \mathbf{p}_5) 
		\,\\ &\times\mathcal{Y}_1^m\left(\frac{\mathbf{p}_4 - \mathbf{p}_5}{2}\right) 
		\, \chi_{45}^{1,-m} \, \phi_{45}^0 \, \omega_{45}^0 
		\, a_4^\dagger(\mathbf{p}_4) \, b_5^\dagger(\mathbf{p}_5),	 
		\end{aligned}
	\end{equation}
	where \(\gamma\) is a dimensionless vacuum production strength parameter and  $ a_4^\dagger(\mathbf{p}_4) \, b_5^\dagger(\mathbf{p}_5)$  create a quark-antiquark pair with momenta $\mathbf{p}_4$ and $\mathbf{p}_5$. \(\mathcal{Y}_1^m(\mathbf{p}) = |\mathbf{p}| Y_1^m(\theta_p, \phi_p)\) is the solid harmonic polynomial reflecting the momentum-space distribution.  \(\omega_{45}^0 \), \(\phi_{45}^0 \) and \(\chi_{45}^{1,-m}\) denote the color, flavor singlets and spin triplet factor of  the quark pair, respectively. These components ensure the created quark pair carries the vacuum quantum number  $J^{PC}=0^{++}$.

	By standard derivation,  the partial-wave  transition amplitude  of \(A \to BC\) can be represented as
 \begin{widetext}
\begin{equation}
	\begin{aligned}
		&\,\,\, M^{SL}_{A \to BC}(\mathbf{P}) = 
		\gamma
		\sqrt{\frac{4\pi(2L+1)}{2J_A+1}}
		\,\sum_{M_{J_B} M_{J_C}}
		\langle L0;\, S(M_{J_B}+M_{J_C}) \,|\, 
		J_A(M_{J_B}+M_{J_C}) \rangle
		\langle J_B M_{J_B}\, J_C M_{J_C} \,|\,
		S(M_{J_B}+M_{J_C}) \rangle
		\\
		&
		\times\Pi_{A,B,C}
		\langle 
		\chi^{12,4}_{S_B M_{S_B}} 
		\chi^{3,5}_{S_C M_{S_C}}
		\,|\,
		\chi^{1,2,3}_{S_A M_{S_A}} 
		\chi^{4,5}_{1-m}
		\rangle
		\langle 
		\phi^{12,4}_B \phi^{3,5}_C 
		\,|\,
		\phi^{1,2,3}_A \phi^{4,5}_0
		\rangle
		\,
		I^{M_{L_A},\,m}_{M_{L_B},\,M_{L_C}}(\mathbf{P})
		\end{aligned}
\end{equation}
with $\mathbf{P}$ representing the three-momentum of particle $B$ in the c.m. frame of particle $A$. 
The $I^{M_A,\,m}_{M_{L_B},\,M_{L_C}}(\mathbf{P})$ are the spatial overlaps
of the initial and final states, which can be written as

\begin{align}
I^{M_A,\,m}_{M_{L_B},\,M_{L_C}}(\mathbf{P})
= \int & d^3\mathbf{p}_1 \, d^3\mathbf{p}_2 \, d^3\mathbf{p}_3
\, d^3\mathbf{p}_4 \, d^3\mathbf{p}_5 \;
\delta^{(3)}(\mathbf{p}_1 + \mathbf{p}_2 + \mathbf{p}_3 - \mathbf{p}_A)
\nonumber 
  \delta^{(3)}(\mathbf{p}_1 + \mathbf{p}_4 + \mathbf{p}_3 - \mathbf{p}_B)
\, \delta^{(3)}(\mathbf{p}_2 + \mathbf{p}_5 - \mathbf{p}_C)
\, \delta^{(3)}(\mathbf{p}_4 + \mathbf{p}_5)
\nonumber \\
& \times \psi^{*}_{n_B L_B M_{L_B}}(\mathbf{p}_1, \mathbf{p}_4, \mathbf{p}_3)
\, \psi^{*}_{n_C L_C M_{L_C}}(\mathbf{p}_2, \mathbf{p}_5)
\, \psi_{n_A L_A M_{L_A}}(\mathbf{p}_1, \mathbf{p}_2, \mathbf{p}_3)
\nonumber  \mathcal{Y}
_{1m}\!\left(\frac{\mathbf{p}_4 - \mathbf{p}_5}{2}\right) ,
\end{align}
 and $\Pi_{A,B,C}$ accounts for the sum over intermediate megnetic indices with the relevant Clebsch–Gordan coefficients as
	\begin{equation}
		\begin{aligned}
			\Pi_{A,B,C}\to
            \sum_{\substack{
					M_{L_A}, M_{S_A} 
					M_{L_B}, M_{S_B} \\ 
					M_{L_C}, M_{S_C}, m
			}}
			&\langle L_A M_{L_A}\, S_A M_{S_A} \,|\,
			J_A (M_{J_B}+M_{J_C}) \rangle\quad\times
			\langle L_B M_{L_B}\, S_B M_{S_B} \,|\, J_B M_{J_B} \rangle
			\,\\
			&\quad\times\langle L_C M_{L_C}\, S_C M_{S_C} \,|\, J_C M_{J_C} \rangle
			\langle 1,m;\, 1,-m \,|\, 0,0 \rangle .
		\end{aligned}
	\end{equation}
It is worth mentioning that here the simple harmonic oscillator (SHO) wave functions are employed in spatial integration,
	\enlargethispage{2\baselineskip} 

		\begin{equation}
			\begin{aligned}
				\psi_{n_{\rho},n_{\lambda}}(l_\rho,m_\rho,l_\lambda,m_\lambda)
				=&\, 3^{3/4}\,(-i)^{2n_{\rho}+l_\rho} 
				\sqrt{\frac{2 n_\rho!}{\Gamma(n_\rho + l_\rho + 3/2)}} \left(\frac{1}{\alpha_\rho}\right)^{3/2+l_\rho} 
				L_{n_\rho}^{\,l_\rho+1/2}\left(\frac{\mathbf{p}_\rho^2}{\alpha_\rho^2}\right) 
				e^{- \mathbf{p}_\rho^2/(2\alpha_\rho^2)} 
				\mathcal{Y}_{l_\rho}^{m_\rho}(\mathbf{p}_\rho)\\[1mm]
				& \times (-i)^{2n_{\lambda}+l_\lambda} 
				\sqrt{\frac{2 n_\lambda!}{\Gamma(n_\lambda + l_\lambda + 3/2)}} 
				\left(\frac{1}{\alpha_\lambda}\right)^{3/2+l_\lambda} 
				L_{n_\lambda}^{\,l_\lambda+1/2}\left(\frac{\mathbf{p}_\lambda^2}{\alpha_\lambda^2}\right) 
				e^{- \mathbf{p}_\lambda^2/(2\alpha_\lambda^2)} 
				\mathcal{Y}_{l_\lambda}^{m_\lambda}(\mathbf{p}_\lambda)
			\end{aligned}.
		\end{equation}	
	\end{widetext}
The harmonic oscillator parameters $\alpha_{\rho/\lambda}$ are determined by matching the effective radius of the full wave function expanded in 15 sets of Gaussians  with that of SHO  wave function, such that
\begin{equation}
	\sum_{i,j=1}^{15}N_{\rm{nor}}\cdot c_j^{*}\,c_i\,
	\langle \phi_j \vert \rho^{2} \vert \phi_i \rangle
	=
	\langle \Psi_{\scriptscriptstyle \rm{SHO}} \vert \rho^{2} \vert 
	\Psi_{\scriptscriptstyle \rm{SHO}} \rangle.
\end{equation}
Consequently, the SHO wave function accurately approximates the spatial extent of the full wave function, and can  be utilized directly in calculating the  coupling vertices. This approach significantly enhances the computational efficiency of the integral evaluation with the $\eta$ function during the pole-searching procedure.

 The momentum $\mathbf{P}$ is the the three momentum of the final-state particle in the initial-state rest frame, given by $| \mathbf{P}(E) | = \lambda(E^2,M_B^2,M_C^2)^{1/2}/(2E)$ where  $\lambda(x,y,z)=x^2+y^2+z^2-2xy-2yz-2xz$ is the K\"{a}ll\'{e}n function.
Then, the coupling vertex function $f_{SL}(E)$ in Eq.~(\ref{etax}), which describes the interaction between $|1\rangle$ and $|E_{i}\rangle$ 
in the extended Lee-Friedrichs scheme, can be obtained as
\begin{equation}
	f_{SL}(E) = \sqrt{\mu\, P(E)} \, M_{SL}\big(P(E)\big),
\end{equation}
 where $\sqrt{\mu P(E)}$ is a phase-space factor and $\mu=({E_{B}E_{C}})/({E_{B}+E_{C}})$ is the reduced mass of the two-body system.
 
   To regulate the high energy behavior of the QPC coupling of the dispersion relation integral, once-subtracted dispersion relation is used in calculating the $\eta$ function as
 \begin{equation}
 	\eta(x)=x-m_{0}-(s_{0}-x)\sum_{i,S,L}\displaystyle\int_{\omega_{th,i}}^{\infty}\dfrac{|f_{SL}(E)|^2}{(x-E)(s_{0}-E)}dE,
 \end{equation}
 where $s_{0}$ is taken at the threshold $\omega_{th,N\pi}$ . In this formulation,  the subtraction constant is implicitly absorbed into the bare mass of the bare $N^*$ state.

 \section{Numerical Analysis}
 \label{sec:results}

The approach differs from the conventional strategy for resonance studies. Usually, the potential model parameters are typically determined by fitting the predicted eigenvalues to the physical mass spectrum, with decay widths subsequently calculated through a decay model in a separate procedure. 
In this study, we used the potential model to generate the bare mass spectrum and the wave functions, which serve as input of bare masses and the coupling vertices in the extended LF scheme. By fitting the resulting pole masses and widths to experimental data of all the $1S$ and $1P$ $N^*$ states, we determined the potential model parameters and the vacuum production strength $\gamma$. Using these determined values, we subsequently calculated the pole position of the $2S$ state, which can be considered a theoretical prediction of this framework .

 The coupled channels chosen in the analysis include  $\pi N$, $\Delta\pi$ and $\eta N$, which are the leading decay modes of 1$P$- and 2$S$-wave nucleons as listed in the PDG table. While an infinite tower of channels with matching quantum numbers could, in principle, couple to a given state, we assume that contributions from higher-mass virtual loops are effectively absorbed into the parameters of the quark potential model or suppressed by the once-subtracted dispersion relation.
  Although including sub-dominant channels is feasible in principle, the resulting proliferation of branch cuts and Riemann sheets would render the numerical identification of poles increasingly inefficient.
 
 In the numerical calculation,  the masses and wave functions of nucleons are obtained by solving the Hamiltonian eigenfunction in the potential model  with the GEM.  To facilitate further calculation of coupling vertices in the LF scheme, we determine an effective oscillator parameter $\alpha_{\rho/\lambda}$  by matching the root-mean-square radius with that of SHO basis.  Consequently, the coupling vertices between the nucleon and the continuum states in the LF scheme  can be expressed in an analytical form,  which significantly improves the efficiency in searching poles on the complex plane.
 
The potential parameters and the vacuum production strength $\gamma$ are determined by matching the predicted complex poles to experimental values. Since these parameters are fixed at the pole level, they implicitly account for coupled-channel dynamics. This optimized parameter set is then applied to the $2S$  state to explore the properties of higher nucleon excitations within the same formalism.

\begin{table}[H]
	\renewcommand\arraystretch{1.3}
	\begin{ruledtabular}
	\centering
		\begin{tabular}{cccc}
			Parameter & Value&Parameter & Value\\
			\hline
			$b$      & 0.10 $\pm$ 0.03 & $\alpha_{SO}$ & 0.61 $\pm$ 0.20 \\
			$\alpha$ & 1.06 $\pm$ 0.22  & $C_{0}$  & $-0.35$ $\pm$ 0.30 \\
             $\sigma$ & 0.66 $\pm$ 0.17 & $\gamma$ & 6.52 $\pm$ 0.49 \\    
	\end{tabular}
	\caption{ Parameters  in potential model and QPC model.}
	\label{tab:parameters}
    \end{ruledtabular}
\end{table}

 Our theoretical framework involves six free parameters. 
 Five of them are associated with the potential model—namely, the confinement parameter $b$, the strong coupling constant $\alpha_{s}$, the smearing parameter  $\sigma$, the spin–orbit coefficient $\alpha_{SO}$, and the zero-point constant  $C_0$.  The sixth  is the dimensionless coupling strength 
$\gamma$, which characterizes the QPC  vertex.  The constituent quark mass of $u/d$ is empirically fixed at 0.35 GeV.
These parameters are determined by fitting the mass of the ground state and the pole positions of the  1P-wave nucleons given by PDG~\cite{ParticleDataGroup:2024cfk}.
The resultant optimal parameter set is summarized in Table~\ref{tab:parameters}.

\twocolumngrid
Using the parameter set, we calculate the mass and pole positions of the ground state of the $1P$- and $2S$-wave nucleons, as listed in Table~\ref{tab:poles}, where  $n$ represents the principal quantum number with $n$ = 1 corresponding to the ground state. To further investigate the dynamical origin of these poles, we trace  their trajectories in relevant physical and un-physical sheets in Fig.~\ref{fig:pole-trajectory}, as the vacuum coupling parameter changes.

The LF model provides an exactly solvable framework that explicitly incorporates the properties of Gamow states, and it allows for a explicit representation of the resonance wave function in RHS, as shown in Eq.~(\ref{resonancewavefunction}).  Beyond the pole positions and trajectories,  the internal nature of a hadronic  state is often characterized by its elementariness~($Z$) and compositeness~($X$). However, some theoretical subtleties regarding their interpretation must be noted. 

For a stable bound state case,  the elementariness $Z$ and compositeness coefficients  $X_{i}$ represents the probability of finding the bare discrete state and the $i$-th continuum state, respectively, within the physical state $|z_B\rangle$ (see Eq.~(\ref{boundstatewavefunction})). 
The elementariness and compositeness  are defined as
\begin{equation}
 	Z=N_{B}^{2}, \qquad
 	X_i = N_{B}^{2} \,
 	\int_{\omega_{{\rm th},i}}^{\infty} \dfrac{ |f_{i}(E)|^2 \, dE}{\displaystyle (z_B - E)^2}.
 \end{equation}
The bound-state energy $z_B$ is real and situated below the lowest threshold, and $Z$ and $X_i$ are both real, satisfying the normalization condition $Z+\sum_i X_i=1$. As such, they admit a standard probability explanation~\cite{Weinberg:1962hj,Hyodo:2013nka,Sekihara:2014kya}.

However, for the resonance state, one can define analogous quantities    $Z'$ and $X_{i}'$ as 
 \begin{equation}
 	Z'= N_{R}^{2}, \qquad
 	X_i' = N_{R}^{2} \,
 	\int_{\omega_{{\rm th},i}}^{\infty} \dfrac{ |f_{i}(E)|^2 \, dE}{\displaystyle [z_R - E]_+^2}.
 \end{equation}
While this definition is formally plausible, the required deformation of integration contour into the unphysical RS renders $Z'$ and $X_{i}'$ complex-valued.  As a result, they no longer possess a direct probability explanation~\cite{Xiao:2016mon}.
{To adopt  a qualitative measure of resonance composition, different methods for evaluating the quantities of resonances were discussed in detail in the Ref.~\cite{Kinugawa:2024crb}.}
 Here, we adopt the following prescription based on the absolute value of these coefficients 
 \begin{equation}
 	\tilde{Z} \equiv 
 	\frac{|Z'|}{\sum_{i} |X'_{i}| + |Z'|},
 		\qquad
 	\tilde{X}_{i} \equiv 
 	\frac{|X'_{i}|}{\sum_{i} |X'_{i}| + |Z'|}.
 	\label{Eqcompositeness}
 \end{equation}
Alternative approaches to define elementariness and compositeness for resonance could be found in literature, such as  
{Ref.~\cite{Guo:2015daa,Aceti:2014ala,Kamiya:2015aea,Kamiya:2016oao}.}
These measures of elementariness and compositeness provide a qualitative means to differentiate bare-state dominance from continuum-driven dynamics. Such a classification sheds light on the formation mechanisms and the internal nature of the poles listed in Table \ref{compositeness}. We will address the properties of these states in the following.

\begin{table*}[t]
\begin{ruledtabular}
\setlength{\tabcolsep}{8pt}
	\renewcommand{\arraystretch}{1.6}
	\label{tab:poles}
	\begin{tabular}{cccccccc}
		Notation 
		& Bare mass 
		& Pole 
		& State 
		& PDG (BW) 
		& PDG (Poles) 
		& Ref.~\cite{Wang:2023snv} 
		& Ref.~\cite{Suzuki:2009nj} \\
		\midrule[0.5pt]
		
		$1\,^{2}P_{1/2^{-}}$ 
		& 1.656 
		& $1.631-0.065i$ 
		& $N(1535)$ 
		& $1.530-0.075i$ 
		& $1.510-0.055i$ 
		& $1.504-0.037i$ 
		& $1.540-0.191i$ \\
		
		$1\,^{2}P_{3/2^{-}}$ 
		& 1.691 
		& $1.528-0.063i$ 
		& $N(1520)$ 
		& $1.515-0.055i$ 
		& $1.510-0.055i$ 
		& $1.482-0.063i$ 
		& $1.521-0.058i$ \\
		
		$1\,^{4}P_{1/2^{-}}$ 
		& 1.621 
		& $1.610-0.043i$ 
		& $N(1650)$ 
		& $1.650-0.063i$ 
		& $1.665-0.067i$ 
		& $1.678-0.064i$ 
		& $1.642-0.041i$ \\
		
		$1\,^{4}P_{3/2^{-}}$ 
		& 1.839 
		& $1.842-0.129i$ 
		& $N(1700)$ 
		& $1.720-0.100i$ 
		& $1.700-0.100i$ 
		& -- 
		& -- \\
		
		$1\,^{4}P_{5/2^{-}}$ 
		& 1.802 
		& $1.677-0.013i$ 
		& $N(1675)$ 
		& $1.675-0.074i$ 
		& $1.655-0.067i$ 
		& $1.652-0.060i$ 
		& $1.654-0.077i$ \\
		
		\midrule
		$2\,^{2}S_{1/2^{+}}$ 
		& 1.799 
		& $1.487-0.070i$ 
		& $N(1440)$ 
		& $1.440-0.175i$ 
		& $1.370-0.085i$ 
		& $1.353-0.102i$ 
		& $1.364-0.105i$ \\
	\end{tabular}
    \caption{Pole positions of selected $N^*$ states. The unit is GeV.}
    \end{ruledtabular}
\end{table*}

\begin{table*}[t]
\begin{ruledtabular}
	\setlength{\tabcolsep}{10pt}
	\renewcommand{\arraystretch}{1.5}
	\label{compositeness}
	\begin{tabular}{ccccccccc}
		State  
		& $Z$ 
		& $\tilde{Z}$ 
		& $X_{\pi N}$ 
		& $\tilde{X}_{\pi N}$ 
		& $X_{\pi \Delta}$ 
		& $\tilde{X}_{\pi \Delta}$ 
		& $X_{\eta N}$ 
		& $\tilde{X}_{\eta N}$ \\
		\midrule[0.5pt]
		$N(1535)$ 
		& {$0.65+0.15i$} 
		& 47.5\% 
		& {$0.08-0.18i$} 
		& 13.7\% 
		& {$0.37+0.14i$} 
		& 27.9\% 
		& {$-0.11-0.11i$} 
		& 10.8\% \\
		
		$N(1520)$ 
		& {$0.74-0.30i$} 
		& 56.3\% 
		& {$0.28+0.08i$} 
		& 20.7\% 
		& {$-0.08+0.25i$} 
		& 18.6\% 
		& {$0.05-0.03i$} 
		& 4.4\% \\
		
		$N(1650)$ 
		& {$ 1.08+0.12i$} 
		& 70.9\% 
		& {$ 0.01-0.07i$} 
		& 4.5\% 
		& {$ 0.13+0.03i$} 
		& 9.0\% 
		& {$ -0.22-0.09i$} 
		& 15.6\% \\
		
		$N(1700)$ 
		& {$0.52+0.19i $} 
		& 50.8\% 
		& {$ -0.00+0.01i$} 
		& 0.7\% 
		& {$0.48-0.21i $} 
		& 47.6\% 
		& {$ 0.00+0.01i$} 
		& 0.9\% \\
		
		$N(1675)$ 
		& {$0.61-0.13i $} 
		& 60.0\% 
		& {$0.02+0.03i $} 
		& 3.3\% 
		& {$ 0.34+0.09i$} 
		& 33.4\% 
		& {$ 0.03+0.01i$} 
		& 3.3\% \\
		
		$N(1440)$ 
		& {$ 0.44-0.25i$} 
		& 39.6\% 
		& {$ 0.11+0.30i$} 
		& 25.0\% 
		& {$ 0.41-0.06i$} 
		& 32.6\% 
		& {$0.04+0.00i $}  
		& 2.8\% \\
	\end{tabular}
    \caption{The compositeness and elementariness of selected $N^*$ states.}  
\end{ruledtabular}
\end{table*}

\subsection{$N(1440)$ $J^P=1/2^+$}

\begin{figure*}
    \centering
    \includegraphics[width=1\linewidth]{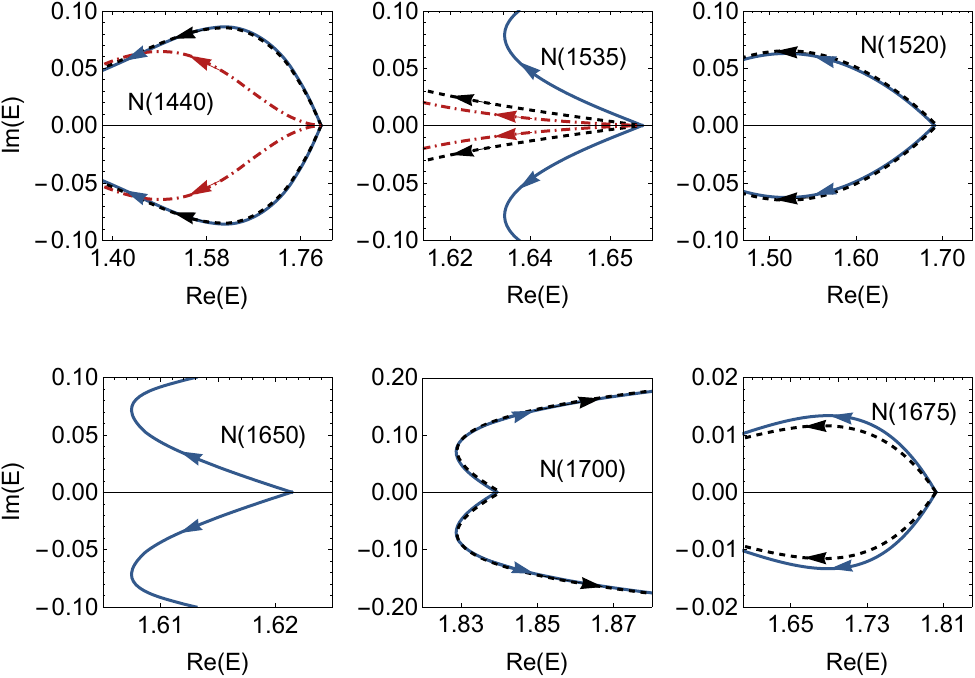}
    \caption{Pole trajectories { with the arrow indicating the moving direction when the coupling strength increases.} The blue solid, black dashed, and red dash-dotted curves represent the pole trajectories on the fourth, third, and second Riemann sheets, respectively. The unit is GeV.)}
    \label{fig:pole-trajectory}
\end{figure*}

Within this framework, using the parameters determined from the 1$P$-wave section, the bare mass of 2$S$ state is predicted to be 1.8 GeV. This value is consistent with typical constituent quark model expectations from the first radial excitation of the nucleon.  However, the physical $N(1440)$ resonance appears at a much lower mass. As illustrated by the pole trajectory in  Fig.~\ref{fig:pole-trajectory},  the inclusion of coupled-channel dynamics leads to significant mass renormalization, {where the arrow indicates that the pole moves downward as the coupling strength increases}. 
As the coupling strength $\gamma$ to the $\pi N$, $\pi\Delta$ and $\eta N$ channels is gradually increased to the fitted value of $6.52$, the $2S$ pole moves drastically downward from its bare position, and eventually resides at  (Re $M_{R}$, -Im $M_{R}$) = (1.487, 0.070)~GeV, which is close the $N$(1440) listed by PDG. This substantial shift of approximately 320 MeV demonstrated that the roper resonance is highly non-perturbative object, where the interaction with the meson-baryon continuum plays a decisive role in forming the physical mass and width.

A long-standing puzzle in baryon spectroscopy is the ``level-inversion" problem, where the $ N$(1440)  is experimentally observed to be lighter than the $N(1520)$ and $N(1535)$ states, a feature that contradicts the conventional quark models. Our results show that the meson-baryon cloud induced by the coupled channel effect might be a key mechanism for resolving the level-inversion difficulty, as the $2S$ state exhibits a much higher sensitivity to the continuum dressing than the $1P$ states.

The internal structure of $N(1440)$ might be further studied by the approximated elementariness $\tilde Z$ and compositeness  $\tilde X_i$  listed in Table ~\ref{compositeness}. Due to the complex nature of these coefficients for resonances, we employ the normalized absolute-value prescription for a qualitative assessment. We find that the bare-state fraction $\tilde Z$ is less than 40$\%$,  close to the result of 36.9\%  reported in Ref.~\cite{Wang:2023snv}. This low value of $\tilde Z$ also indicate that the Roper resonance is not a pure three-quark state but is instead predominantly composite. Physically, it can be viewed as a bare three-quark core surrounded by a dense meson-baryon cloud, with the $\pi N$ and $\pi \Delta$
components  provide dominant compositeness. This picture is consistent with recent analyses from Refs.~\cite{Mokeev:2025hhe,Suzuki:2009nj}.
However, the dynamical model in Ref.~\cite{Suzuki:2009nj} suggests that, due to strong coupled-channel effects, the same bare state  also induces a relatively broad resonance around 1.8 GeV, potentially  associated with the $N(1710)$. In our theoretical framework, we have not found the signal of such a higher-mass resonance. This difference might be attributed to the specific form of the vertex functions or the channel space. It suggests that the high-energy behavior of the transition potentials or the interference between different partial waves might contribute to additional shadow pole structure.

\subsection{$N(1535)$ and $N(1650) $ $J^P=1/2^-$}

The internal nature of the $N(1535)$ resonance has long been a subject of  theoretical debate. Within the chiral unitary approach with a natural renormalization scheme, Ref.~\cite{HyodoJidoHosaka2008}  argued that the $N(1535)$ resonance contains a significant genuine quark component, distinguishing it from states like $\Lambda(1405)$ which are considered to be predominantly dynamically generated. This result is supported by the  J\"ulich-Bonn
 dynamical coupled-channel approach~\cite{Ronchen:2022hqk}.
However, the degree of its ``elementariness" remains a point of contention.

Alternatively, other studies based on chiral effective other studies based on SU(3) chiral effective Lagrangian theories suggest that 
$N(1535)$ can be generated primarily through meson–baryon dynamics, without the need of introducing an explicit s-channel bare state, but rather as a consequence of the non-perturbative resummation of coupled-channel interactions.
In these approaches, the strong couplings to the $K\Lambda$ and $K\Sigma$ channels are found to play a particularly important role in the formation of the resonance, which indicates that the possibility of the 
$N(1535)$ being a dynamically generated state cannot be ruled out~\cite{Kaiser:1995cy,Nieves:2001wt,Bruns:2010sv,Doring:2009uc}.

In our LF framework,  by introducing a bare state $1 {}^2P_{1/2^{-}}$, we find  a resonance in the complex plane at  (Re $M_{R}$, -Im $M_{R}$) = (1.631, 0.063) GeV. This result, while slightly higher than the PDG average, si remarkably consistent with the  result of Hamiltonian eﬀective field theory in Ref.~\cite{Liu:2015ktc}, which gives  (Re $M_{R}$, -Im $M_{R}$) = (1.602, 0.089) GeV. The discrepancy with PDG values may arise from the simplified treatment of high-energy tails of the vertex functions. Nevertheless, our compositeness analysis in  Table~\ref{compositeness} provide a clear structural distinction: the $N(1535)$ exhibits a significantly larger compositeness compared to the  $N(1650)$ , which remains relatively compact. This conclusion aligns with the spectral density analysis in Ref.~\cite{Wang:2023snv}.

We also observe that that $N(1535)$ and $N(1650)$ are nearly degenerate in our calculation.
According to  the PDG~\cite{ParticleDataGroup:2024cfk}, these two states not only share the same quantum numbers $J^P=1/2^-$ but also exhibit similar dominant decay channels, particularly into the $\pi N$ and $\eta N$ channels. It is therefore physically motivated to interpret these two states as the physical manifestations of the mixing between the $1 {}^2 P_{1/2^{-}}$ and $1 {}^4 P_{1/2^{-}}$ bare configurations. The mixing angle   can be extracted from the photoproduction data (often reported as $25^\circ \sim 30^\circ$), such as  $\gamma p \rightarrow \eta p$~\cite{He:2008ty} and $\gamma p\rightarrow \pi^{0}p$~\cite{Zhong:2011ti,Xiao:2015gra}.

In the limit of exact SU(6) symmetry, these two states would be degenerate. However, the inclusion of spin-dependent interactions breaks the symmetry, inducing mixing between the two states.
In Ref.~\cite{PhysRevD.110.116034}, a comparable mixing angle of approximately $\theta=26^\circ$ can also be obtained by
introducing OGE and OBE potentials and considering the mixing between these two states via the tensor interaction.
The mixing mechanism  can also provide a phenomenological explanation  for the distinct branching fractions of $N(1535)$ and $N(1650)$ into $\pi N$ and $\eta N$ channels~\cite{PhysRevD.110.116034}.

In the present frameword, by phenomenologically adopting this mixing angle $\theta=26^\circ$, the positions of these two poles will shift to (Re $M_{R}$, -Im $M_{R}$) = (1.649, 0.074) GeV and (Re $M_{R}$, -Im $M_{R}$) = (1.603, 0.041) GeV, respectively. It is evident that they appear to split further, suggesting that the experimentally observed states are more preferably  understood as  linear combinations of multiple underlying states. This splitting also indicates that the dynamical coupling to meson-baryon continua and the internal spin-flavor mixing are inextricably linked, both contributing to the observed mass spectrum and decay patterns of the $S_{11}$ nucleon resonances.

\twocolumngrid
\subsection{$N(1520)$ and $N(1700)$ $J^P=3/2^-$}
In the traditional quark model,  the $N(1520)$ and $N(1700)$ are classified as the $1P$-wave excited nucleons with $J^P=3/2^-$, corresponding to the $1^2P_{3/2^-}$ and $1^4P_{3/2^-}$ configurations, respectively. Based on the qualitative decomposition as introduced in Eq.~\eqref{Eqcompositeness}, our calculations yield bare-state fractions  of approximately 50\% for both states. These results suggest that the $N(1520)$ and $N(1700)$ state exhibits a significant composite nature, where meson–baryon components plays a non-negligible role in their formation. This findings for $N(1520)$ is consistent with the spectral density analysis presented in Ref.~\cite{Wang:2023snv}.

A particularly compelling piece of evidence for our model's validity is found in the channel-specific compositeness values $\tilde X_i$. We find that $N(1520)$ is dominated by a large $\pi N$ component, whereas $N(1700)$ is characterized by a predominant $\pi \Delta$ compositeness.  This aligns remarkably well with the PDG branch fractions~\cite{ParticleDataGroup:2024cfk}, which indicate that $N(1520)$ decays primarily to the $\pi N$ channel~($55\sim65$\%) while  $N(1700)$ is known to have a significant decay width into the $\pi \Delta$ channel. This agreement reinforces the physical relevance of the estimated composition, which suggests that each bare state preferentially dresses itself  with specific meson-baryon clouds that reflect its underlining spin-flavor structure.
 
Differ from the other states discussed here, $N(1700)$ is the only 3-star  broad resonance in the PDG, which has larger uncertainties in pole positions.	
In this LF scheme, $N(1700)$ appears as a robust and stable resonance together with the $N(1520)$, despite its weak $\pi N$ (d-wave) signals and large experimental uncertainties.	
Due to the strong coupling of the $1^4P_{3/2^-}$ bare state to the continua, its pole moves from the real axis to the complex energy plane and then shift rapidly upwards as the coupling constant $\gamma$ increases, eventually appearing as a broad resonance pole centered around 1800 MeV with a width of approximately 260 MeV.

Intriguingly,  as the coupling increases, besides the emergence of  $N(1700)$, a dynamically generated state arises from a distant region, eventually settles on the complex (Re $M_{R}$, -Im $M_{R}$) = ($1.468,0.236$) GeV in the third Riemann sheet. Since this pole is located on the complex energy plane when the coupling constant $\gamma$ tends to zero, it implies the molecule origin of this state, that means this pole is predominantly generated by the meson–baryon dynamics rather than a bare quark core.
However, its extremely large width~($\Gamma>470$ MeV) may explain why
this dynamical generated state may bring challenges to its experimental observation due to the feature of $\Gamma>300$ MeV.
This analogous mechanism, two pole structure, frequently occurs in meson systems, such as  $f_{0}(500)$ and   
and $K_{0}^{*}(700)$ \cite{Zhou:2020moj}.
It is noteworthy  that  a similar dynamical generated state found  (Re $M_{R}$, -Im $M_{R}$) = (1.467, 0.083) GeV, corresponding to $N(1520)$ in Ref.~\cite{Garzon:2013pad}.
This state has a comparable mass with ours but a significantly narrower width.

Similarly, for these two states sharing the same quantum numbers, incorporation  a more refined three-body spin-orbit potential, as detailed  in Ref.~\cite{PhysRevD.110.116034}, allows for a direct computation of  the mixing angle, yielding $\theta=20^\circ$. 
By adopting this mixing angle as input, 
 due to the strong coupling of the $^4P_{ 3/2^-}$ state, the pole positions after mixing shift significantly,  locating at  (Re $M_{R}$, -Im $M_{R}$) = (1.466, 0.098) GeV and (Re $M_{R}$, -Im $M_{R}$) = (1.714, 0.073) GeV, corresponding to $N(1520)$ and
$N(1700)$, respectively.
The new position of $N(1520)$ is close to the result in Ref.~\cite{Garzon:2013pad}, while the pole position of $N(1700)$ shows improved consistency  with the PDG averaged values.

\subsection{$N(1675)$ $J^P=5/2^-$}
For the  $N(1675)$ resonance state, the calculated pole width is somewhat narrower than experimental value reported by the PDG. 
This discrepancy likely originates from several factors. First, the transition couplings derived from the  QPC model may carry inherent uncertainties, which directly affect the imaginary of the self energy. Second, our current calculation only incorporates the three most dominant decay channels. Other channels, such as $\rho N$, $\omega N$ and $K\Sigma$, though smaller in branching fractions, may vary the pole position.

According to  Ref.~\cite{Wang:2023snv}, more channels were considered, and the composition of each configuration was analyzed.
Their results indicate that the $\rho N$
 channel contributes approximately 10\% for this state.
 When multiple decay channels are included, as three channels in the present case, the   pole trajectories generally exhibit more complicated behavior.
In particular, interference effects among different channels can reduce the total decay probability, which may lead to a narrower pole width compared with the single-channel case.
The complex amplitudes from various channels can partially cancel each other in the imaginary part of the pole position, reflecting the underlying multichannel dynamics rather than a simple sum of individual widths.
Instead of pursuing an exact agreement with the experimental width, we  focuses more on the pole trajectories.

\section{Summary}
\label{Summary}
In this work, we have investigated the pole structure and Gamow-state compositions of  low-lying nucleon resonances within an extended LF scheme. By integrating the bare states of selected resonance and their coupling to meson-baryon continuum states described by the QPC model, we constructed a multi-channel framework including the $\pi N,\pi\Delta$ and $\eta N$ continua. These three dominant channels are selected to capture the essential dynamics of the system. 
The model parameters  are calibrated by matching the pole masses and widths with those values of the $1P$-wave resonances documented by PDG.

Our framework successfully reproduces six resonances  are successfully reproduced, $N(1535) 1/2^{-}$, $N(1520) 3/2^{-}$, $N(1650) 1/2^{-}$, $N(1700) 3/2^{-}$, $N(1675) 5/2^{-}$ and $N(1440) 1/2^{+}$.
By tracing the pole trajectories  and analyzing their elementariness and compositeness, we demonstrate that the strong coupling  between the $2S$ state and these channels contributes a significant mass shift, and then generating a broad resonance around 1.4 GeV.
Specifically, the elementariness of $N(1440)$ is found to be less than 40\%, which suggests that it is not a pure valence quark state but rather a three-quark core  heavily dressed by baryon-meson clouds. This implies that incorporating unquenched effects into the traditional quark model is promising in understanding the internal structures of excited nucleons and resolving the long-standing mass-inversion puzzle.
In addition, a dynamically generated $J^P=3/2^{-}$ state with a pole position at (Re $M_{R}$, -Im $M_{R}$) = (1.468, 0.236) GeV. This state  is likely to appear accompanied with $N(1700)$, though its  large width likely poses a challenge for clear experimental  identification.

Despite these successes, we acknowledge that  certain uncertainties remains between our pole positions and  and  the PDG values. 
These uncertainties might stem from from three primary aspects:
Firstly, the QPC model employed for the coupling vertex functions represents is a leading-order approximation;
Moreover, the  values for the masses and widths of these resonances listed by PDG are averages with sizable uncertainties, which may propagate into our parameter determination; Lastly, some higher-order coupled channels were omitted to maintain computational tractability.

This study represents our first application of the extended LF framework to the complex dynamics of the baryon system. The results demonstrate the feasibility and robustness of this approach in describing the unquenched baryon spectrum. Further work will focus on refine the vertex functions and extending this systematic investigation to other baryon sections.

\section*{Acknowledgement} Helpful discussion with Zhi-Guang Xiao are appreciated. This work is supported by China National Natural Science Foundation under contract Nos. 12375132, 11975075 and 12175065.

\bibliographystyle{unsrt}  
\bibliography{refs}

\begin{thebibliography}{10}

\bibitem{Roper:1964zza}
L.~David Roper.
\newblock {Evidence for a P-11 Pion-Nucleon Resonance at 556 MeV}.
\newblock {\em Phys. Rev. Lett.}, 12:340--342, 1964.

\bibitem{Suzuki:2009nj}
N.~Suzuki, B.~{Juli{\'a}-D{\'i}az}, H.~Kamano, T.-S.~H. Lee, A.~Matsuyama, and
  T.~Sato.
\newblock Disentangling the {{Dynamical Origin}} of {{P}} 11 {{Nucleon
  Resonances}}.
\newblock {\em Physical Review Letters}, 104(4):042302, 2010.

\bibitem{Liu:2016uzk}
Zhan-Wei Liu, Waseem Kamleh, Derek~B. Leinweber, Finn~M. Stokes, Anthony~W.
  Thomas, and Jia-Jun Wu.
\newblock {Hamiltonian effective field theory study of the ${N^*(1440)}$
  resonance in lattice QCD}.
\newblock {\em Phys. Rev. D}, 95(3):034034, 2017.

\bibitem{Qin:2019hgk}
Si-xue Qin, Craig~D Roberts, and Sebastian~M Schmidt.
\newblock {Spectrum of light- and heavy-baryons}.
\newblock {\em Few Body Syst.}, 60(2):26, 2019.

\bibitem{Wu:2017qve}
Jia-jun Wu, Derek~B. Leinweber, Zhan-wei Liu, and Anthony~W. Thomas.
\newblock {Structure of the Roper Resonance from Lattice QCD Constraints}.
\newblock {\em Phys. Rev. D}, 97(9):094509, 2018.

\bibitem{Lang:2016hnn}
C.~B. Lang, L.~Leskovec, M.~Padmanath, and S.~Prelovsek.
\newblock {Pion-nucleon scattering in the Roper channel from lattice QCD}.
\newblock {\em Phys. Rev. D}, 95(1):014510, 2017.

\bibitem{Suenaga:2022ajn}
Daiki Suenaga and Atsushi Hosaka.
\newblock {Decays of Roper-like singly heavy baryons in a chiral model}.
\newblock {\em Phys. Rev. D}, 105(7):074036, 2022.

\bibitem{Clement:2020dby}
H.~Clement, T.~Skorodko, and E.~Doroshkevich.
\newblock {Possibility of dibaryon formation near the N*(1440)N threshold:
  Reexamination of isoscalar single-pion production}.
\newblock {\em Phys. Rev. C}, 106(6):065204, 2022.

\bibitem{Ball:1967zzb}
James~S. Ball, Gordon~L. Shaw, and David~Y. Wong.
\newblock {Two-Channel Model of P-11pi-N Partial-Wave Amplitude}.
\newblock {\em Phys. Rev.}, 155:1725--1727, 1967.

\bibitem{Zou:2025nnw}
Bing-Song Zou.
\newblock {Roper resonance $N^*(1440)$ from charmonium decays}.
\newblock 9 2025.

\bibitem{Zhao:2006an}
Qiang Zhao and Frank~E. Close.
\newblock {Quarks, diquarks and QCD mixing in the N* resonance spectrum}.
\newblock {\em Phys. Rev. D}, 74:094014, 2006.

\bibitem{Long:2011rt}
Bingwei Long and U.~van Kolck.
\newblock {The Role of the Roper in Chiral Perturbation Theory}.
\newblock {\em Nucl. Phys. A}, 870-871:72--82, 2011.

\bibitem{Chen:2017pse}
Chen Chen, Bruno El-Bennich, Craig~D. Roberts, Sebastian~M. Schmidt, Jorge
  Segovia, and Shaolong Wan.
\newblock {Structure of the nucleon{\textquoteright}s low-lying excitations}.
\newblock {\em Phys. Rev. D}, 97(3):034016, 2018.

\bibitem{Tan:2025kjk}
Yue Tan, Zi-Xuan Ma, Xiaoyun Chen, Xiaohuang Hu, Youchang Yang, Qi~Huang, and
  Jialun Ping.
\newblock {Investigating the nature of N(1535) and {\ensuremath{\Lambda}}(1405)
  in a quenched chiral quark model}.
\newblock {\em Phys. Rev. D}, 111(9):096018, 2025.

\bibitem{Cheng:2025sdp}
Peng Cheng, Langtian Liu, Ya~Lu, and Craig~D. Roberts.
\newblock {Insights into Nucleon Resonances via Continuum Schwinger Function
  Methods}.
\newblock 12 2025.

\bibitem{Godfrey:1985xj}
S.~Godfrey and Nathan Isgur.
\newblock {Mesons in a Relativized Quark Model with Chromodynamics}.
\newblock {\em Phys. Rev. D}, 32:189--231, 1985.

\bibitem{Capstick:1986ter}
Simon Capstick and Nathan Isgur.
\newblock {Baryons in a relativized quark model with chromodynamics}.
\newblock {\em Phys. Rev. D}, 34(9):2809--2835, 1986.

\bibitem{Glozman:1995fu}
L.~Ya. Glozman and D.~O. Riska.
\newblock {The Spectrum of the nucleons and the strange hyperons and chiral
  dynamics}.
\newblock {\em Phys. Rept.}, 268:263--303, 1996.

\bibitem{PhysRevD.110.116034}
Hui-Hua Zhong, Ming-Sheng Liu, Ru-Hui Ni, Mu-Yang Chen, Xian-Hui Zhong, and
  Qiang Zhao.
\newblock Unified study of nucleon and $\mathrm{\ensuremath{\Delta}}$ baryon
  spectra and their strong decays with chiral dynamics.
\newblock {\em Phys. Rev. D}, 110:116034, Dec 2024.

\bibitem{Julia-Diaz:2006odw}
B.~Julia-Diaz and D.~O. Riska.
\newblock {The Role of qqqq anti-q components in the nucleon and the N(1440)
  resonance}.
\newblock {\em Nucl. Phys. A}, 780:175--186, 2006.

\bibitem{ParticleDataGroup:2024cfk}
S.~Navas et~al.
\newblock {Review of particle physics}.
\newblock {\em Phys. Rev. D}, 110(3):030001, 2024.

\bibitem{Bijker:2015gyk}
R.~Bijker, J.~Ferretti, G.~Galat{\`a}, H.~Garc{\'\i}a-Tecocoatzi, and
  E.~Santopinto.
\newblock {Strong decays of baryons and missing resonances}.
\newblock {\em Phys. Rev. D}, 94(7):074040, 2016.

\bibitem{Hunt:2018wqz}
B.~C. Hunt and D.~M. Manley.
\newblock {Updated determination of $N^*$ resonance parameters using a unitary,
  multichannel formalism}.
\newblock {\em Phys. Rev. C}, 99(5):055205, 2019.

\bibitem{Gegelia:2016xcw}
Jambul Gegelia, Ulf-G. Mei{\ss}ner, and De-Liang Yao.
\newblock {The width of the Roper resonance in baryon chiral perturbation
  theory}.
\newblock {\em Phys. Lett. B}, 760:736--741, 2016.

\bibitem{Burkert:2017djo}
Volker~D. Burkert and Craig~D. Roberts.
\newblock {Colloquium : Roper resonance: Toward a solution to the fifty year
  puzzle}.
\newblock {\em Rev. Mod. Phys.}, 91(1):011003, 2019.

\bibitem{Shklyar:2012js}
V.~Shklyar, H.~Lenske, and U.~Mosel.
\newblock {{\ensuremath{\eta}}-meson production in the resonance-energy
  region}.
\newblock {\em Phys. Rev. C}, 87(1):015201, 2013.

\bibitem{Hoferichter:2023mgy}
Martin Hoferichter, Jacobo~Ruiz de~Elvira, Bastian Kubis, and Ulf-G.
  Mei{\ss}ner.
\newblock {Nucleon resonance parameters from Roy{\textendash}Steiner
  equations}.
\newblock {\em Phys. Lett. B}, 853:138698, 2024.

\bibitem{Ronchen:2022hqk}
Deborah R{\"o}nchen, Michael D{\"o}ring, Ulf-G. Mei{\ss}ner, and Chao-Wei Shen.
\newblock {Light baryon resonances from a coupled-channel study including
  $\mathbf {K\Sigma }$ photoproduction}.
\newblock {\em Eur. Phys. J. A}, 58(11):229, 2022.

\bibitem{Anisovich:2011fc}
A.~V. Anisovich, R.~Beck, E.~Klempt, V.~A. Nikonov, A.~V. Sarantsev, and
  U.~Thoma.
\newblock {Properties of baryon resonances from a multichannel partial wave
  analysis}.
\newblock {\em Eur. Phys. J. A}, 48:15, 2012.

\bibitem{Arndt:2006bf}
R.~A. Arndt, W.~J. Briscoe, I.~I. Strakovsky, and R.~L. Workman.
\newblock {Extended partial-wave analysis of piN scattering data}.
\newblock {\em Phys. Rev. C}, 74:045205, 2006.

\bibitem{Pearce:1990uj}
B.~C. Pearce and B.~K. Jennings.
\newblock {A Relativistic meson exchange model of pion - nucleon scattering}.
\newblock {\em Nucl. Phys. A}, 528:655--675, 1991.

\bibitem{Wang:2023snv}
Yu-Fei Wang, Ulf-G. Mei{\ss}ner, Deborah R{\"o}nchen, and Chao-Wei Shen.
\newblock Examination of the nature of the {{N}} * and {{$\Delta$}} resonances
  via coupled-channels dynamics.
\newblock {\em Physical Review C}, 109(1):015202, January 2024.

\bibitem{Krehl:1999km}
O.~Krehl, C.~Hanhart, S.~Krewald, and J.~Speth.
\newblock What is the structure of the {{Roper}} resonance?
\newblock {\em Physical Review C}, 62(2):025207, 2000.

\bibitem{Xiao:2015gra}
Li-Ye Xiao, Xu~Cao, and Xian-Hui Zhong.
\newblock {Neutral pion photoproduction on the nucleon in a chiral quark
  model}.
\newblock {\em Phys. Rev. C}, 92(3):035202, 2015.

\bibitem{Xiao:2016mon}
Zhiguang Xiao and Zhi-Yong Zhou.
\newblock {Partial Wave Decomposition in Friedrichs Model With Self-interacting
  Continua}.
\newblock {\em J. Math. Phys.}, 58:072102, 2017.

\bibitem{Xiao:2016dsx}
Zhiguang Xiao and Zhi-Yong Zhou.
\newblock Virtual states and the generalized completeness relation in the
  {{Friedrichs}} model.
\newblock {\em Physical Review D}, 94(7):076006, October 2016.

\bibitem{Xiao:2023lpv}
Zhiguang Xiao and Zhi-Yong Zhou.
\newblock {On the generalized Friedrichs-Lee model with multiple discrete and
  continuous states*}.
\newblock {\em Chin. Phys.}, 49(8):083102, 2025.

\bibitem{Gadella:2004}
O.~Civitarese and M.~Gadella.
\newblock Physical and mathematical aspects of gamow states.
\newblock {\em Communications on Pure and Applied Mathematics}, 396(2):41--113,
  2004.

\bibitem{Lee:1954iq}
T.~D. Lee.
\newblock {Some Special Examples in Renormalizable Field Theory}.
\newblock {\em Phys. Rev.}, 95:1329--1334, 1954.

\bibitem{Friedrichs:1948}
K.~O. Friedrichs.
\newblock On the perturbation of continuous spectra.
\newblock {\em Commun. Pure Appl. Math.}, 1(4):361--406, 1948.

\bibitem{Isgur:1978xj}
Nathan Isgur and Gabriel Karl.
\newblock {P Wave Baryons in the Quark Model}.
\newblock {\em Phys. Rev. D}, 18:4187, 1978.

\bibitem{Isgur:1978wd}
Nathan Isgur and Gabriel Karl.
\newblock {Positive Parity Excited Baryons in a Quark Model with Hyperfine
  Interactions}.
\newblock {\em Phys. Rev. D}, 19:2653, 1979.

\bibitem{Bhaduri:1981pn}
R.~K. Bhaduri, L.~E. Cohler, and Y.~Nogami.
\newblock A unified potential for mesons and baryons.
\newblock {\em Il Nuovo Cimento A (1965-1970)}, 65(3):376--390, 1981.

\bibitem{Kalman:1988br}
C.~S. Kalman and B.~Tran.
\newblock Baryon spectrum in a potential quark model.
\newblock 102(3):835--879.

\bibitem{Liu:2019wdr}
Ming-Sheng Liu, Kai-Lei Wang, Qi-Fang L{\"u}, and Xian-Hui Zhong.
\newblock {$\Omega$ baryon spectrum and their decays in a constituent quark
  model}.
\newblock {\em Phys. Rev. D}, 101(1):016002, 2020.

\bibitem{Pervin:2007wa}
Muslema Pervin and Winston Roberts.
\newblock {Strangeness -2 and -3 baryons in a constituent quark model}.
\newblock {\em Phys. Rev. C}, 77:025202, 2008.

\bibitem{Roberts:2007ni}
W.~Roberts and Muslema Pervin.
\newblock {Heavy baryons in a quark model}.
\newblock {\em Int. J. Mod. Phys. A}, 23:2817--2860, 2008.

\bibitem{Hiyama:2003cu}
E.~Hiyama, Y.~Kino, and M.~Kamimura.
\newblock {Gaussian expansion method for few-body systems}.
\newblock {\em Prog. Part. Nucl. Phys.}, 51:223--307, 2003.

\bibitem{Micu:1968mk}
L.~Micu.
\newblock {Decay rates of meson resonances in a quark model}.
\newblock {\em Nucl. Phys. B}, 10:521--526, 1969.

\bibitem{Carlitz:1970xb}
Robert~D. Carlitz and M.~Kislinger.
\newblock {Regge amplitude arising from su(6)w vertices}.
\newblock {\em Phys. Rev. D}, 2:336--342, 1970.

\bibitem{LeYaouanc:1972vsx}
A.~Le~Yaouanc, L.~Oliver, O.~Pene, and J.~C. Raynal.
\newblock {Naive quark pair creation model of strong interaction vertices}.
\newblock {\em Phys. Rev. D}, 8:2223--2234, 1973.

\bibitem{Zhou:2017dwj}
Zhi-Yong Zhou and Zhiguang Xiao.
\newblock {Understanding $X(3862)$, $X(3872)$, and $X(3930)$ in a
  Friedrichs-model-like scheme}.
\newblock {\em Phys. Rev. D}, 96(5):054031, 2017.
\newblock [Erratum: Phys.Rev.D 96, 099905 (2017)].

\bibitem{Weinberg:1962hj}
Steven Weinberg.
\newblock {Elementary particle theory of composite particles}.
\newblock {\em Phys. Rev.}, 130:776--783, 1963.

\bibitem{Hyodo:2013nka}
Tetsuo Hyodo.
\newblock {Structure and compositeness of hadron resonances}.
\newblock {\em Int. J. Mod. Phys. A}, 28:1330045, 2013.

\bibitem{Sekihara:2014kya}
Takayasu Sekihara, Tetsuo Hyodo, and Daisuke Jido.
\newblock {Comprehensive analysis of the wave function of a hadronic resonance
  and its compositeness}.
\newblock {\em PTEP}, 2015:063D04, 2015.

\bibitem{Kinugawa:2024crb}
Tomona Kinugawa and Tetsuo Hyodo.
\newblock {Compositeness of hadrons, nuclei, and atomic systems}.
\newblock {\em Eur. Phys. J. A}, 61(7):154, 2025.

\bibitem{Guo:2015daa}
Zhi-Hui Guo and J.~A. Oller.
\newblock {Probabilistic interpretation of compositeness relation for
  resonances}.
\newblock {\em Phys. Rev. D}, 93(9):096001, 2016.

\bibitem{Aceti:2014ala}
F.~Aceti, L.~R. Dai, L.~S. Geng, E.~Oset, and Y.~Zhang.
\newblock {Meson-baryon components in the states of the baryon decuplet}.
\newblock {\em Eur. Phys. J. A}, 50:57, 2014.

\bibitem{Kamiya:2015aea}
Yuki Kamiya and Tetsuo Hyodo.
\newblock {Structure of near-threshold quasibound states}.
\newblock {\em Phys. Rev. C}, 93(3):035203, 2016.

\bibitem{Kamiya:2016oao}
Yuki Kamiya and Tetsuo Hyodo.
\newblock {Generalized weak-binding relations of compositeness in effective
  field theory}.
\newblock {\em PTEP}, 2017(2):023D02, 2017.

\bibitem{Mokeev:2025hhe}
V.~I. Mokeev and D.~S. Carman.
\newblock {Roper Resonance Structure and Exploration of Emergent Hadron Mass
  from CLAS Electroproduction Data}.
\newblock 11 2025.

\bibitem{HyodoJidoHosaka2008}
Daisuke~Jido Tetsuo~Hyodo and Atsushi Hosaka.
\newblock Origin of resonances in the chiral unitary approach.
\newblock {\em Prog. Part. Nucl. Phys.}, 2008.

\bibitem{Kaiser:1995cy}
Norbert Kaiser, P.~B. Siegel, and W.~Weise.
\newblock {Chiral dynamics and the S11 (1535) nucleon resonance}.
\newblock {\em Phys. Lett. B}, 362:23--28, 1995.

\bibitem{Nieves:2001wt}
J.~Nieves and E.~Ruiz~Arriola.
\newblock {The S(11) - N(1535) and - N(1650) resonances in meson baryon
  unitarized coupled channel chiral perturbation theory}.
\newblock {\em Phys. Rev. D}, 64:116008, 2001.

\bibitem{Bruns:2010sv}
Peter~C. Bruns, Maxim Mai, and Ulf~G. Meissner.
\newblock {Chiral dynamics of the S11(1535) and S11(1650) resonances
  revisited}.
\newblock {\em Phys. Lett. B}, 697:254--259, 2011.

\bibitem{Doring:2009uc}
M.~Doring and K.~Nakayama.
\newblock {The Phase and pole structure of the N*(1535) in pi N
  ---{\ensuremath{>}} pi N and gamma N ---{\ensuremath{>}} pi N}.
\newblock {\em Eur. Phys. J. A}, 43:83--105, 2010.

\bibitem{Liu:2015ktc}
Zhan-Wei Liu, Waseem Kamleh, Derek~B. Leinweber, Finn~M. Stokes, Anthony~W.
  Thomas, and Jia-Jun Wu.
\newblock {Hamiltonian effective field theory study of the $\mathbf{N^*(1535)}$
  resonance in lattice QCD}.
\newblock {\em Phys. Rev. Lett.}, 116(8):082004, 2016.

\bibitem{He:2008ty}
Jun He, B.~Saghai, and Zhenping Li.
\newblock {Study of $\eta$ photoproduction on the proton in a chiral
  constituent quark approach via one-gluon-exchange model}.
\newblock {\em Phys. Rev. C}, 78:035204, 2008.

\bibitem{Zhong:2011ti}
Xian-Hui Zhong and Qiang Zhao.
\newblock {$\eta$ photoproduction on the quasi-free nucleons in the chiral
  quark model}.
\newblock {\em Phys. Rev. C}, 84:045207, 2011.

\bibitem{Zhou:2020moj}
Zhi-Yong Zhou and Zhiguang Xiao.
\newblock {Two-pole structures in a relativistic Friedrichs{\textendash}Lee-QPC
  scheme}.
\newblock {\em Eur. Phys. J. C}, 81(6):551, 2021.

\bibitem{Garzon:2013pad}
E.~J. Garzon, J.~J. Xie, and E.~Oset.
\newblock {Case in favor of the $N^*(1700)(3/2^-)$}.
\newblock {\em Phys. Rev. C}, 87(5):055204, 2013.

\end{thebibliography}

\end{document}